\documentclass[aps,prl,floatfix,twocolumn,showpacs,amsmath,amssymb]{revtex4}
\usepackage{graphicx}
\usepackage{dcolumn}
\usepackage{bm}
\begin{document}

\title{Spin-liquid versus spiral-order phases in the anisotropic triangular 
       lattice}
\author{Luca F. Tocchio,$^{1}$ H\'el\`ene Feldner,$^{1}$ Federico Becca,$^{2}$
Roser Valent\'i,$^{1}$ and Claudius Gros$^{1}$}
\affiliation{
$^{1}$Institute for Theoretical Physics, 
       University of Frankfurt, 
       Max-von-Laue-Stra{\ss}e 1, D-60438 Frankfurt a.M., Germany \\
$^{2}$CNR-IOM-Democritos National Simulation Centre 
       and International School for Advanced Studies (SISSA), 
       Via Bonomea 265, I-34136, Trieste, Italy
            }

\date{\today} 

\begin{abstract}
We study the competition between magnetic and spin-liquid phases in the Hubbard
model on the anisotropic triangular lattice, which is described by two hopping 
parameters $t$ and $t^\prime$ in different spatial directions and is relevant 
for layered organic charge-transfer salts. By using a variational approach that
includes spiral magnetic order, we provide solid evidence that a spin-liquid 
phase is stabilized in the strongly-correlated regime and close to the 
isotropic limit $t^\prime/t=1$. Otherwise, a magnetically ordered spiral state
is found, connecting the (collinear) N\'eel and the (coplanar) $120^\circ$ 
phases. The pitch vector of the spiral phase obtained from the unrestricted 
Hartree-Fock approximation is substantially renormalized in presence of 
electronic correlations, and the N\'eel phase is stabilized in a wide regime 
of the phase diagram, i.e., for $t^\prime/t < 0.75$. We discuss these results 
in the context of organic charge-transfer salts.
\end{abstract}

\pacs{71.10.Fd, 71.27.+a, 75.10.-b}

\maketitle

\emph{Introduction}-- The combined presence of strong electron 
interaction and geometrical frustration leads to a plethora of 
interesting phenomena, like superconductivity, 
metal-insulator (Mott) transition, or purely quantum
paramagnets, the so-called spin liquids. In this context, 
the organic charge-transfer salts $\kappa$-(ET)$_2$X~\cite{def1} 
play an important role.~\cite{kanoda,powell,tsunetsugu} A large variety of 
phases have been found when changing temperature, pressure or 
the nature of the anion X, ranging from correlated metals with 
superconductivity at low temperatures, to insulators with 
magnetic order.~\cite{kanoda2,elsinger,lefebvre,limelette}
Even more interestingly, a metal-insulator transition to a 
pure non-magnetic Mott insulating state has been detected for 
the compound with X=Cu$_2$(CN)$_3$.~\cite{kanoda3,manna} Recently, 
another family of organic materials, denoted by 
Et$_n$Me$_{4-n}Pn$[Pd(dmit)$_2$]$_2$,~\cite{def2} has been shown 
to display different quantum phases, including valence-bond solid 
and spin-liquid states.~\cite{yamaura,shimizu} From a quantum 
chemical perspective, the simplest possible effective Hamiltonian 
for the organic charge-transfer salts is the Hubbard model 
(after an appropriate particle-hole transformation) on the 
anisotropic triangular lattice at half filling. 
Indeed, in these materials, strongly dimerized organic 
molecules are arranged in stacked two-dimensional triangular 
lattices; each dimer has (on the average) a charge state 
with one hole, implying a half-filled conducting band.
In addition, a sizable effective Coulomb repulsion is 
felt by two holes on the same dimer, while longer range 
correlations are much smaller.~\cite{valenti,nakamura}

\begin{figure}
\includegraphics[height=0.8\columnwidth,angle=270]{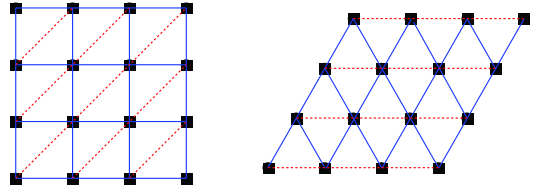}
\caption{\label{fig:lattice}
Illustration of the anisotropic triangular lattice in the square topology 
(left), used in this work, and in the equivalent triangular topology (right). 
Solid and dashed lines denote hopping amplitudes $t$ and $t^\prime$, 
respectively.}
\end{figure}

The Hubbard model is defined by:
\begin{equation}\label{eq:hubbard}
{\cal H}=-\sum_{i,j,\sigma} t_{ij} c^\dagger_{i,\sigma} c_{j,\sigma} + 
\textrm{h.c.} + U \sum_{i} n_{i,\uparrow} n_{i,\downarrow},
\end{equation}
where $c^\dagger_{i,\sigma} (c_{i,\sigma})$ creates (destroys) an electron 
with spin $\sigma$ on site $i$, $n_{i,\sigma}=c^\dagger_{i,\sigma}c_{i,\sigma}$
is the electronic density, $t_{ij}$ is the hopping amplitude and $U$ is the 
on-site Coulomb repulsion. In this work, we focus our attention on the 
half-filled case, where the number of electrons $N_e$ equals the number of 
sites $L$, and consider a square lattice with a nearest-neighbor hopping $t$, 
along $(1,0)$ and $(0,1)$ directions, and a further next-nearest-neighbor 
hopping $t^\prime$ along $(1,1)$; this choice of the hopping amplitudes is 
topologically equivalent to the anisotropic triangular lattice, see 
Fig.~\ref{fig:lattice}. According to recent density functional theory 
calculations,~\cite{valenti,nakamura,scriven,nakamura2} the ratio $t^\prime/t$ 
appropriate for organic salts lies in the range $[0.3,1.3]$, with the 
spin-liquid compound $\kappa$-(ET)$_2$Cu$_2$(CN)$_3$ located at 
$t^\prime/t \simeq 0.83$. 

A major issue in the Hubbard model on the anisotropic triangular lattice 
is the possibility of stabilizing a spin-liquid phase, compatible with the 
experimental data. From one side, several approaches have proposed the 
existence of a spin liquid-region for $t^\prime/t<1$, based on path-integral 
renormalization group (PIRG),~\cite{imada} dynamical mean-field theory 
(DMFT),~\cite{kyung} exact diagonalization,~\cite{clay} and variational Monte 
Carlo (VMC).~\cite{tocchio} In addition, several studies suggested a possible 
spin-liquid phase even for the isotropic case $t^\prime/t=1$, close to the 
metal-insulator transition.~\cite{mila,antipov,senechal,kawakami}
From the other side, for generic values of the ratio $t^\prime/t$, magnetic 
states with incommensurate order may be expected and indeed have been proposed
by different mean-field approaches, like for instance within the Hartree-Fock
(HF) approximation~\cite{krish,lacroix,inui} or the renormalized mean-field 
method.~\cite{powell2} However, due to the difficulty of constructing 
{\it correlated} magnetic states with {\it generic} ordering vectors, none 
of the previous studies was able to perform a fair comparison between 
spin-liquid and spiral states. In this respect, some progress to deal with
incommensurate magnetism in the Heisenberg model has been done by means of
analytic approximations,~\cite{mckenzie,thomale,hauke} or density-matrix 
renormalization group (DMRG) calculations,~\cite{white} even if here the 
long-range nature of the magnetic correlations is not addressed. Instead, 
incommensurate correlations in the Hubbard model have been only marginally 
addressed by Cluster-DMFT.~\cite{ohashi}

In order to go beyond the previous studies, we approach 
this problem by implementing correlated variational wave 
functions which describe magnetic states with generic 
incommensurate order. This can be achieved by starting
from the spiral states obtained at the HF level and including,
in a second step, many body correlations. In this way, we are 
able to treat incommensurate spiral order and non-magnetic 
states on the same level and determine which state is stabilized 
for a given value of frustration $t^\prime/t$ and Coulomb repulsion
$U$. Variational approaches may contain, as a matter of principle, 
a bias towards ordered states. However, we showed 
previously~\cite{tocchio,tocchio2} that very accurate 
results are obtained in a wide regime of the 
parameters when using, as in the present work, generalized
Gutzwiller wave functions with long-range Jastrow correlations and
backflow corrections.

Here, we confirm that a spin-liquid phase is favored for 
$t^\prime/t \simeq 0.85$, while magnetic spiral order becomes 
competitive close to the isotropic point, i.e., for 
$t^\prime/t \simeq 0.95$, and for $t^\prime/t \simeq 0.75-0.8$. 
In addition, we explicitly study the effect of correlations on a 
mean-field state showing how the energy and the pitch vector of 
the spiral state are modified when electronic correlations are 
taken into account by the VMC method.       

\begin{figure}
\includegraphics[width=\columnwidth]{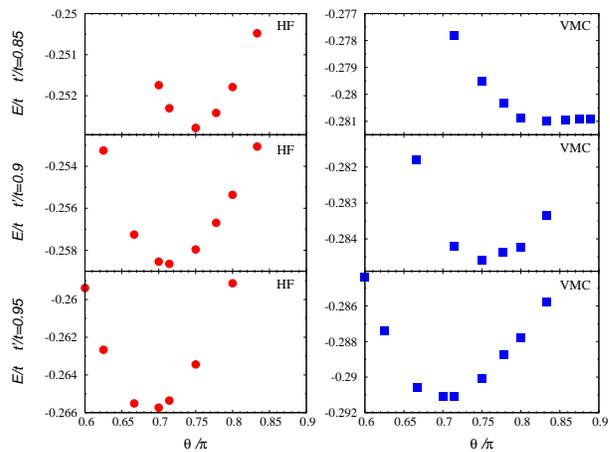}
\caption{\label{fig:AngleMin}
(Color online) HF (left) and VMC (right) energies of the spiral state, 
for $U/t=16$, as a function of the pitch angle $\theta$ (in unit of $\pi$), 
for $t^\prime/t=0.85$ (top), $0.9$ (middle) and $0.95$ (bottom). Cluster of 
sizes $l\times l$ have been used, with $l=12,\,14,\,16,\,18,\,20$.}
\end{figure}

\emph{Unrestricted Hartree Fock}--
The unrestricted HF state is obtained by performing
a mean-field decoupling of Eq.~(\ref{eq:hubbard}):
\begin{eqnarray}
\label{eq:HMF}
H_{\textrm{HF}}&=&
-\sum_{i,j,\sigma} t_{ij} c^\dagger_{i,\sigma} c_{j,\sigma} + \textrm{h.c.} \\
&& + U \sum_i \left[ \langle n_{i,\downarrow} \rangle n_{i,\uparrow}
+ \langle n_{i,\uparrow} \rangle n_{i,\downarrow}\right] \nonumber \\
&& - U \sum_i \left[ 
\langle c^\dagger_{i,\uparrow} c_{i,\downarrow} \rangle 
c^\dagger_{i,\downarrow} c_{i,\uparrow} 
+ \langle c^\dagger_{i,\downarrow} c_{i,\uparrow} \rangle 
c^\dagger_{i,\uparrow} c_{i,\downarrow} \right] \nonumber \\
&& - U \sum_i \left[ \langle n_{i,\uparrow} \rangle 
\langle n_{i,\downarrow} \rangle - 
\langle c^\dagger_{i,\uparrow} c_{i,\downarrow} \rangle 
\langle c^\dagger_{i,\downarrow} c_{i,\uparrow} \rangle \right],
\nonumber
\end{eqnarray}
which contains $4L$ independent mean-field parameters 
to be computed self-consistently:
$\langle n_{i,\uparrow} \rangle$, $\langle n_{i,\downarrow} \rangle$,
$\langle c^\dagger_{i,\uparrow} c_{i,\downarrow} \rangle$, and
$\langle c^\dagger_{i,\downarrow} c_{i,\uparrow} \rangle$ for each site. 
Here, we slightly restrict the variational freedom 
and impose the spin order to be coplanar in the $x{-}y$ plane, 
namely, we look for solutions with
$\langle n_{i,\uparrow} \rangle = \langle n_{i,\downarrow} \rangle$,
thus reducing the number of independent parameters to $3L$. The mean-field 
Hamiltonian~(\ref{eq:HMF}) can be diagonalized and the ground state 
$|\textrm{SP}\rangle$ can be computed by filling the lowest-energy 
single-particle orbitals. 

Here, we are interested in describing the nature of the insulating state, 
which is stabilized for sufficiently large on-site interactions. In this
regime, the optimal HF solutions display a spiral magnetic order, 
which, for $t^\prime/t\le 1$, may be parametrized through a single pitch 
angle $\theta\in[2\pi/3,\pi]$. Indeed, nearest-neighbor spins, along 
$(1,0)$ and $(0,1)$ directions, form an angle $\theta$, while 
next-nearest-neighbor spins, along the $(1,1)$ direction, form an angle 
$2\theta$; a pitch angle of $\theta=\pi$ corresponds to N\'eel order, 
suitable for $t^\prime=0$, and $\theta=2\pi/3$ to the $120^\circ$ order, 
suitable for $t^\prime=t$. On finite-size clusters with periodic-boundary
conditions, only the set of commensurate pitch angles is accessible; 
for $L=l \times l$, the allowed values are $\theta=2\pi n/l$, 
with $n$ being an integer. 

\emph{Variational Monte Carlo}--
Within the VMC approach, we construct magnetic states with spiral order by
applying correlation terms on top of spiral states. Since the optimal pitch
angle in the presence of electron correlations may differ from the one obtained
at the HF level, several different values of $\theta$ are considered in the 
VMC calculations. We employ a spin-spin Jastrow factor to correctly describe
fluctuations orthogonal to the plane where the magnetic order lies, i.e., 
${\cal J}_s=\exp [1/2 \sum_{i,j} u_{i,j} S_i^z S_j^z ]$.~\cite{becca}
A further density-density Jastrow factor
${\cal J}_c=\exp [1/2 \sum_{i,j} v_{i,j} n_i n_j ]$ 
(that includes the on-site Gutzwiller term $v_{i,i}$) is considered to adjust 
electron correlations. All the $u_{i,j}$'s and the $v_{i,j}$'s are optimized 
for every independent distance $|i-j|$. The correlated state is then given by 
$|\Psi_{\textrm{SP}}\rangle = {\cal J}_s {\cal J}_c |\textrm{SP}\rangle$.

\begin{figure}
\includegraphics[width=\columnwidth]{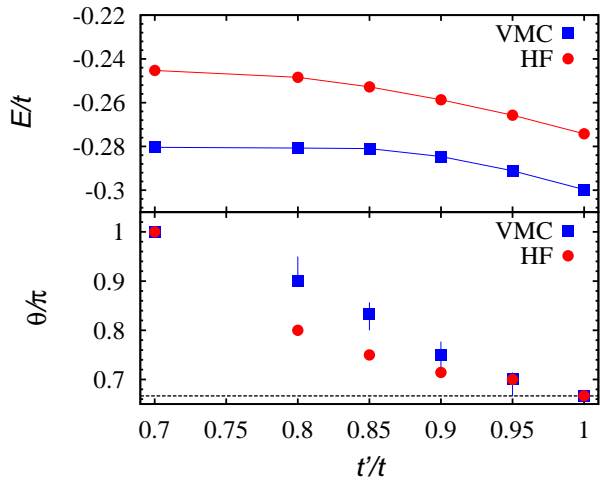}
\caption{\label{fig:HFvsVMC}
(Color online) Upper panel: HF (red circles) and VMC (blue squares) energies 
of the optimal spiral state as a function of $t^\prime/t$ for $U/t=16$. 
Lower panel: the pitch angle $\theta$ (in unit of $\pi$) of the optimal spiral
state as a function of $t^\prime/t$ for $U/t=16$. The dotted horizontal line 
corresponds to $\theta=2\pi/3$, suitable for the isotropic point 
$t^\prime/t=1$. The error bars in the VMC calculations are due to the 
finite-size limitations in the accessible pitch angles.}
\end{figure}

In order to describe a non-magnetic insulator we construct, in a first step, 
an uncorrelated wave function given by the ground state $|\textrm{BCS}\rangle$
of a superconducting BCS Hamiltonian:~\cite{gros,zhang,gros2,edegger}
\begin{equation}\label{eq:meanfield}
{\cal H}_{\rm{BCS}} = \sum_{k,\sigma} \xi_k 
c^\dagger_{k,\sigma} c_{k,\sigma} + \sum_{k} \Delta_k 
c^\dagger_{k,\uparrow} c^{\dagger}_{-k,\downarrow} + \textrm{h.c.},
\end{equation}
where both the free-band dispersion $\xi_k$ and the pairing amplitudes 
$\Delta_k$ are variational functions. We use the parametrization
\begin{eqnarray}
\label{eq:epsilon}
\xi_k &=& -2\tilde{t}(\cos k_x+\cos k_y)
-2\tilde{t}^\prime\cos (k_x + k_y)-\mu, \\
\Delta_k &=& 2\Delta_{\textrm{BCS}}(\cos k_x-\cos k_y),
\label{eq:Delta}
\end{eqnarray}
where the effective hopping amplitude $\tilde{t}^\prime$, the effective 
chemical potential $\mu$, and the pairing field $\Delta_{\textrm{BCS}}$ are 
variational parameters to be optimized. The $d$-wave symmetry of the pairing 
function introduced in Eq.~(\ref{eq:Delta}) is found to be the best 
variational state in all the range $t'/t\le 1$. The correlated state 
$|\Psi_{\textrm{BCS}}\rangle={\cal J}_c |\textrm{BCS}\rangle$ 
allows then to describe a non-magnetic Mott insulator for 
a sufficiently singular Jastrow factor $v_q\sim 1/q^2$ 
($v_q$ being the Fourier transform of $v_{i,j}$).~\cite{capello} 

\begin{figure}
\includegraphics[width=\columnwidth]{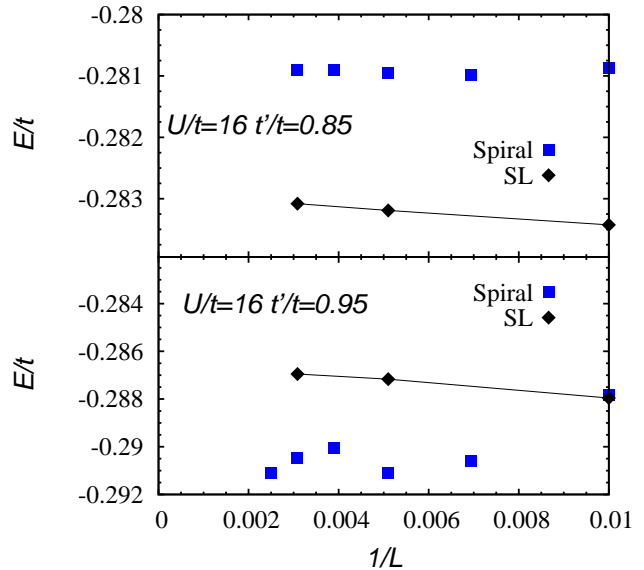}
\caption{\label{fig:spiral_SL}
(Color online) Upper panel: VMC energy of the optimal spiral (blue squares) 
and spin-liquid (black diamonds) states as a function of the inverse system 
size $1/L$, with $L$ ranging from $10 \times 10$ to $20 \times 20$. Data 
refer to the case $t^\prime/t=0.85$ and $U/t=16$. Lower panel: the same as 
in the upper panel but for $t^\prime/t=0.95$.}
\end{figure}

A size-consistent and efficient way to further improve the correlated states 
$|\Psi_{\textrm{BCS}}\rangle$ and $|\Psi_{\textrm{SP}}\rangle$ is based on 
backflow correlations. In this approach, each orbital that defines the 
unprojected states $|\textrm{BCS}\rangle$ and $|\textrm{SP}\rangle$ is taken 
to depend upon the many-body configuration, in order to incorporate virtual 
hopping processes.~\cite{tocchio2} All results presented here are obtained by 
fully incorporating the backflow corrections and optimizing individually every
variational parameter in $\xi_k$ and $\Delta_k$, in the Jastrow factors 
${\cal J}_c$ and ${\cal J}_s$, as well as in the backflow corrections. 

Finally, we want to mention that the $d$-wave symmetry of the pairing
function, introduced in Eq.~(\ref{eq:Delta}), is in agreement with previous VMC
studies~\cite{nandini,watanabe} carried out using variational wave functions
not containing long-range Jastrow and backflow correlations.

\emph{Evolution of the pitch angle}-- 
In Fig.~\ref{fig:AngleMin}, the energy per site of the 
spiral state is presented as a function of the pitch angle $\theta$,
for different values of the frustrating hoppings $t^\prime/t$ and $U/t=16$,
both for the HF and the VMC calculations. We mention that the pitch angle 
is only weakly dependent on $U/t$ in the insulating region. 
For finite lattices the set of allowed pitch angles is determined by
commensurability, we therefore include in Fig.~\ref{fig:AngleMin}
results for several cluster sizes. The overall behavior of
the energy per site versus $\theta$ is smooth, indicating that
size effects are under control. In the region of 
$0.8 \lesssim t^\prime/t \lesssim 0.9$, a very shallow energy 
landscape is observed in VMC, while, for larger values of the 
frustrating hopping, the minimum is much more pronounced.

We find that Jastrow and backflow terms influence 
the periodicity of the spiral order and that the inclusion of the 
correlation factors induce a sizable gain in the energy per site, 
strongly improving the quality of the variational state. 
In Fig.~\ref{fig:HFvsVMC}, we present the evolution of the energy 
and of the pitch angle $\theta$ of the optimal spiral state as 
a function of $t^\prime/t$, for $U/t=16$. In the intermediate range of 
frustration $0.75 \lesssim t^\prime/t \lesssim 0.9$, the explicit treatment 
of electronic correlations (within the VMC level) renormalizes the angle of 
the HF optimal spiral state and values much closer to $\pi$ are stabilized.
As a result, correlation effects stabilize the N\'eel phase 
(i.e., $\theta=\pi$) in a wider regime, e.g., for $t^\prime/t < 0.75$. 
Close to the isotropic point, i.e., for $t^\prime/t \lesssim 1.0$, the optimal
pitch angle shifts rapidly towards $2\pi/3$. 

\emph{Spiral vs. spin-liquid state}-- Let us now move to the main result of 
the present work and compare the optimal spiral and spin-liquid states. 
We find that for $t^\prime/t=0.85$ and $U/t=16$, which are suitable parameters
for $\kappa$-(ET)$_2$Cu$_2$(CN)$_3$, the lowest variational energy is achieved
by a magnetically disordered state, see Fig.~\ref{fig:spiral_SL}. The results
are only weakly dependent on the cluster size. On the other hand, for a larger
value of the frustrating hopping, i.e., $t^\prime/t=0.95$, a spiral state 
with angle $\theta/\pi \simeq 0.7$ is favored over the spin liquid. 
In this case, although slightly larger size effects are present for the spiral
state, the trend is clear. 

\begin{figure}
\includegraphics[width=\columnwidth]{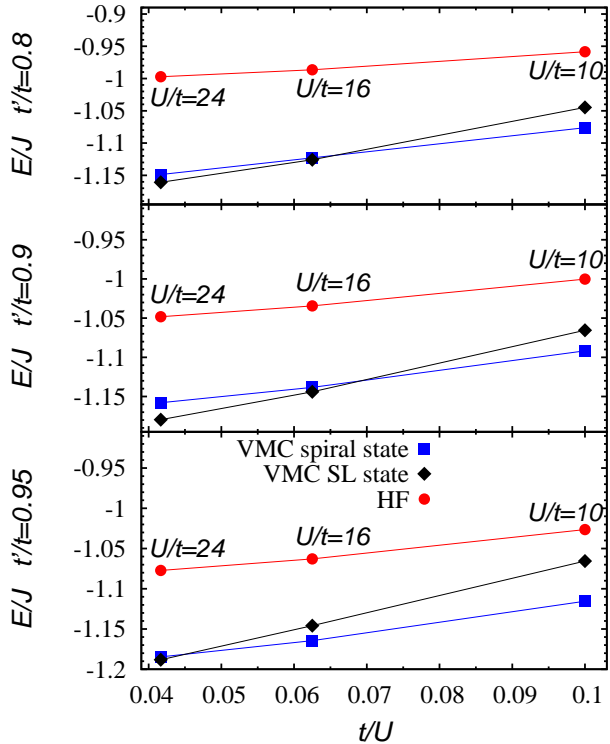}
\caption{\label{fig:Jscale}
(Color online) The energy (in unit of $J=4t^2/U$) as a function of $t/U$ for 
the HF approximation (red circles), the optimal spiral state in VMC 
(blue squares), and the spin-liquid state (black diamonds) for
$t^\prime/t=0.8$ (top), $0.9$ (middle) and $0.95$ (bottom).}
\end{figure}

\begin{figure}
\includegraphics[width=0.9\columnwidth]{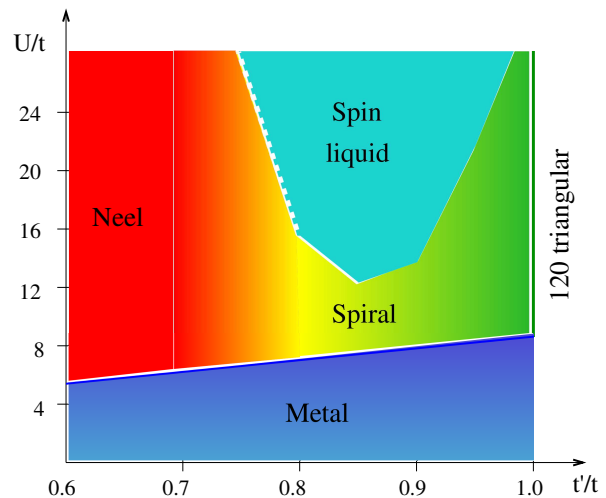}
\caption{\label{fig:diagram}
(Color online) Schematic phase diagram of the Hubbard model on the anisotropic
triangular lattice, as obtained by VMC: metal (blue), insulator with magnetic
N\'eel order (red), insulator with spiral magnetic order (gradient 
red-yellow-green), and spin liquid (cyan). The $120^\circ$ ordered state with 
$\theta= 2\pi/3$ (vertical green line) is stable only for $t^\prime/t=1$. The 
spiral state illustrated with the red-yellow gradient has a pitch angle 
$\theta$ ranging from $\pi$ to $0.9\pi$. The border between the spin-liquid 
and the magnetic state for $t^\prime/t <0.8$ (dashed cyan-white line) is only 
inferred since pitch angles too close to $\pi$ could not be resolved.}
\end{figure}

In Fig.~\ref{fig:Jscale}, we present the VMC energies (in unit of $J=4t^2/U$) 
for the optimal spiral and spin-liquid states as a function of $U/t$ 
and different values of $t^\prime/t$. We find that there is a critical 
$U/t$ above which the spin-liquid state is energetically favored, 
while for smaller values of $U/t$ the magnetically ordered state 
is stabilized. The simple HF energy of the spiral state is also 
reported for comparison; however, the accuracy of the HF state is not 
sufficient to study the competition with the spin-liquid state.
 
\emph{Phase diagram}-- In Fig.~\ref{fig:diagram}, we present 
the final phase diagram. We identify a metallic phase, which is likely 
not superconductive,~\cite{tocchio,dayal} and three insulating 
phases: a phase with commensurate N\'eel order 
(i.e., $\theta=\pi$), a spiral-order phase with
$2\pi/3<\theta<\pi$, and a spin-liquid region. Note that the 
$120^\circ$ order with $\theta=2\pi/3$ is stable only at 
the isotropic point $t^\prime/t=1$. Within the region 
$0.7\lesssim t^\prime/t \lesssim 0.8$ the pitch angle is 
close to $\pi$, and our numerical calculations applied
to finite-size clusters cannot resolve the actual value 
of the optimal $\theta$. Our results indicate that the 
spin-liquid state is not stable asymptotically close to 
the isotropic point, for $t^\prime\to t$, at least not
for any finite $U/t$. We also point out that the spin liquid 
described by our variational wave function is gapless at the 
nodal points $k=(\pm \pi/2,\pm \pi/2)$, see Eq.~(\ref{eq:Delta}). 
Finally, we would like to mention
the fact that, within our present approach, the transition 
from a spiral state to the spin liquid is always first order.
However, we cannot exclude the existence of a continuous 
transition in the exact ground state. In this regard, 
it would be possible to determine the nature of the 
transition by considering BCS-spiral uncorrelated states 
$|\textrm{BCS},\textrm{SP}\rangle$, which are however 
technically extremely demanding, and beyond the scope 
of the present study.

\emph{Conclusions}--
By using a state-of-the-art variational approach, we studied the insulating 
phase of the half-filled Hubbard model on anisotropic triangular lattices with 
$t^\prime/t \le 1$. Through a combined HF and VMC approach, we showed that 
spiral states, with non-trivial pitch angles, are stable in the strongly 
frustrated region. For larger values of interaction a spin-liquid phase 
emerges. These results open two intriguing possibilities. On one side, the 
spin-liquid and the spiral phases may be considered two competing phases. 
In this case, the transition would be expected to be of first order. 
On the other side, the possibility of a new route towards a 
non-magnetic correlated state emerges, namely the spin liquid may be 
considered as an instability emerging from a strongly correlated 
spiral phase. In this case, a second order transition would be expected. 
To resolve this question one would need to consider strongly-correlated 
combined BCS-spiral states, which are however technically very demanding
and left for future studies. Finally, in the parameter region of relevance
for the reported spin-liquid behavior in organic charge-transfer salts,
we also find the spin-liquid phase to be the most stable. Regarding spiral
order, to our knowledge no experimental evidence for this phase has been 
reported for organic charge-transfer salts. It would be desirable to search 
for such a state, especially in the context of new orderings observed in 
these materials.~\cite{lunkenheimer} 

L.F.T. and H.F. acknowledge the support of the German Science Foundation 
through the grant SFB/TRR49.


\begin{thebibliography}{99}
\bibitem{def1} ET=BEDT-TTF stands for bis(ethylenedithio)-tetrathiafulvalene
    and X is a monovalent anion.
\bibitem{def2} Me stands for methyl, Et stands for ethyl,
  Pn denotes a pnictogen atom and dmit is 1,3-dithiole-2-thione-4,5-dithiolate.
\end{thebibliography}
\end{document}